\documentclass[twocolumn]{autart1}    

\usepackage{graphicx}          

\usepackage{amsfonts,amssymb}
\usepackage{dsfont}
\usepackage{mathrsfs}

\usepackage{amsmath} 

\usepackage{mathtools}

\usepackage{color}
\usepackage[dvips]{epsfig}
\usepackage{mathtools}
\usepackage{thmtools}
\usepackage{url}
\usepackage{pgfplots}

\makeatletter
\long\def\@maketablecaption#1#2{\@tablecaptionsize
  \setbox\@tempboxa\hbox{#1. #2}
  \ifdim \wd\@tempboxa >\hsize              
    \unhbox\@tempboxa\par                   
  \else                                     
    \global \@minipagefalse
    \hbox to\hsize{\hfil\box\@tempboxa\hfil}
  \fi}
\makeatother

\newtheorem{assumption}{Assumption}
\newtheorem{lemma}{Lemma}
\newtheorem{theorem}{Theorem}
\newtheorem{definition}{Definition}

\newcommand{\ve}[1]{\boldsymbol{#1}}
\newcommand{\amin}[1]{\underset{#1}{\operatorname{arg\,min}}}

\begin{document}

\begin{frontmatter}

\title{On Asymptotic Analysis of the Two-Stage Approach: Towards Data-Driven Parameter Estimation} 

\author{Braghadeesh Lakshminarayanan}\ead{blak@kth.se},    
\author{Cristian R. Rojas}\ead{crro@kth.se}      

\address{Division of Decision and Control Systems, KTH Royal Institute of Technology, Stockholm, Sweden}  

\begin{keyword}                           
Parameter estimation, System identification, Statistical decision theory, Large sample theory. 
\end{keyword}                             

\begin{abstract}                          
{
In this paper, we analyze the asymptotic properties of the Two-Stage (TS) estimator---a simulation-based parameter estimation method that constructs estimators offline from synthetic data. While TS offers significant computational advantages compared to standard approaches to estimation, its statistical properties have not been previously analyzed in the literature. Under simple assumptions, we establish that the TS estimator is strongly consistent and asymptotically normal, providing the first theoretical guarantees for this class of estimators.}
\end{abstract}

\end{frontmatter}

\section{Introduction}
{
Due to recent advancements in the construction of high-fidelity simulators (or \emph{digital twins})~\cite{vanderhorn2021digital}, it is possible to estimate unknown parameters by generating synthetic input-output data~\cite{piga2024syntheticdatagenerationidentification} under different parameter configurations of such simulators. The resulting datasets, which pair simulated observations with corresponding parameter values, can be used to train a supervised learning model that maps real measurements to parameter estimates. This paved the way to \emph{simulation-driven estimators}~\cite{Diskin,garatti2013new,GHOSH2024111327}, which are constructed offline using synthetic data and later applied to real measurements with minimal computational overhead. Related works leveraging synthetic data offline in identification tasks other than parameter estimation are comprehensively covered in~\cite{piga2024syntheticdatagenerationidentification,forgione2023system}. 

A particularly simple and flexible simulation-based method is the \emph{Two-Stage (TS) estimator}~\cite{garatti2008estimation,garatti2013new}. The TS approach proceeds in two steps: (i) the simulated data is first compressed via a feature map that retains key information in the data, and (ii) a regression model is trained to predict the parameter values from the compressed features. Once trained, the learned estimator is applied directly to real observations to produce a parameter estimate. This framework avoids the need for an explicit likelihood or prediction error function and is especially attractive when inference speed is critical. It supports a wide range of compression functions and regression models, and has been shown to perform competitively against methods such as Prediction-Error Method (PEM), Indirect Inference (II), Particle filters (PF) and Kalman filtering (KF)~\cite{garatti2013new}. Extensions include minimax variants~\cite{lakshminarayanan2023minimax}, statistical decision theoretic interpretations~\cite{BLCRCDC2022}, and applications to controller tuning~\cite{dettu2024data}. Despite the computational advantages of the TS approach, its statistical properties remain mostly unexplored. In particular, no general results are available regarding its consistency or asymptotic distribution. This paper aims to fill this gap. Subject to specific choices of the first and second stages of the TS approach, our main contributions are:
\vspace{-0.5 em}
\begin{itemize}
    \item We demonstrate that the TS estimator converges almost surely to a limit functional as the number of training parameter samples and observation length grow large, thus establishing its strong consistency.
    \item {As a preliminary result, of independent interest, we establish the consistency of sample quantiles as estimators of the quantiles of a distribution}.
    \item We establish the asymptotic normality of the TS approach, characterizing its asymptotic covariance.
\end{itemize}

The rest of the paper is organized as follows: In Section~\ref{subsec: motivation}, we provide a brief motivation for the TS estimator. In Section~\ref{sec: problem statement} we define the problem setup and provide an overview of the TS estimator. Section~\ref{sec: Asymptotic analysis} provides the asymptotic analysis of the TS estimator under simplified assumptions, and a numerical validation of the asymptotic results is provided in Section~\ref{sec: numerical validation}. Finally, we conclude the paper in Section~\ref{sec: Conclusion}.

\textbf{Notation}:
Vectors and matrices are written in bold.
The transpose of a vector $\ve{x}$ is $\ve{x}^{\top}$.
The Gaussian distribution of mean $\mu$ and variance $\sigma^2$ is written $\mathcal{N}(\mu, \sigma^2)$, and the composition of two functions $f$ and $g$ is denoted $g \circ f$.
The order statistics corresponding to a sample vector $\ve{y}^{(k)}$ are collected in the vector $\ve{y}_{\text{order}}^{(k)} = (y_{(1)}^{({k})}, \ldots, y_{(N)}^{({k})})^{\top}$, where $y_{(i)}^{(\ve{\theta})} \leq y_{(j)}^{(\ve{\theta})}$ for every $i \leq j$.
The indicator function of a set $A$ is denoted $\mathds{1}_A\{\cdot\}$, the floor function is written $\lfloor x \rfloor$, and the Jacobian of a function $\ve{J}$ evaluated at a point $\ve{x}$ is $\nabla \ve{J}(\boldsymbol{x})$.

Before formally defining the TS estimator, we discuss its practical advantages. 

\subsection{Motivation} \label{subsec: motivation}
The practical appeal of TS is \emph{fast inference}: once trained offline, the learned mapping returns a parameter estimate in a single pass, whereas traditional point estimation methods (such as PEM and EKF) solve an optimization or filtering problem once real data is collected. To illustrate this, Appendix~\ref{app:comparison} reports a controlled–system study (adapted from~\cite{garatti2013new}) comparing TS with EKF and PEM under identical conditions. TS attains competitive accuracy and the \emph{shortest} inference times; the accuracy setup follows~\cite{garatti2013new}, while the inference–time benchmark is new and included solely to highlight TS’s speed.

}

{
\section{Overview of the TS Estimator} \label{sec: problem statement}
In this section, we define the parameter estimation problem and introduce the Bayes TS estimator as a method to address it, leveraging synthetic data from a high-fidelity simulator to estimate the unknown parameters of a system.

\subsection{Problem setup} \label{subsec: problem_setup}
Consider a high-fidelity simulator $\mathcal{M}(\boldsymbol{\theta})$ of a system, parameterized by $\boldsymbol{\theta} \in \Theta \subseteq \mathbb{R}^d$, $d \geq 1$, which can produce observations $\boldsymbol{y}_N^{(\boldsymbol{\theta})} = (y_1^{(\boldsymbol{\theta})}, \ldots, y_N^{(\boldsymbol{\theta})})^{\top} \in \mathbb{R}^N$ for any $N$. The `true' system is described by $\mathcal{M}(\boldsymbol{\theta}_0)$, where $\boldsymbol{\theta}_0 \in \Theta$ is unknown. Then, the goal of parameter estimation is to determine the parameter $\boldsymbol{\theta}_0$ of the system from observed data $\boldsymbol{y}_N^{(0)} = (y_1^{(0)}, \ldots, y_N^{(0)})^{\top} \in \mathbb{R}^N$. The parameters $\boldsymbol{\theta}$ and observations $\boldsymbol{y}_N^{(\boldsymbol{\theta})}$ are random variables defined on a probability space $(\Omega, \mathcal{F}, \mathbb{P})$. The joint distribution of $\ve{\theta}$ and $\ve{y}_N^{(\ve{\theta})}$ is $\mathbb{P}(\boldsymbol{\theta}, \boldsymbol{y}_N^{(\boldsymbol{\theta})}) = \mathbb{P}_{\boldsymbol{\theta}}(\boldsymbol{y}_N^{(\boldsymbol{\theta})}) \pi(\boldsymbol{\theta})$, where $\pi(\boldsymbol{\theta})$ is a prior distribution of $\boldsymbol{\theta}$ over $\Theta$, and $\mathbb{P}_{\boldsymbol{\theta}}(\boldsymbol{y}_N^{(\boldsymbol{\theta})})$ is the conditional distribution of $\boldsymbol{y}_N^{(\boldsymbol{\theta})}$ given $\boldsymbol{\theta}$. In the case of independent and identically distributed (i.i.d.) data, we have that $y_i^{(\boldsymbol{\theta})} \stackrel{\mathrm{i.i.d.}}{\sim} {f_{\boldsymbol{\theta}}}$, with \(
\mathbb{P}_{\boldsymbol{\theta}}(\boldsymbol{y}_N^{(\boldsymbol{\theta})}) = \prod_{i=1}^N f_{\boldsymbol{\theta}}\bigl(y_i^{(\boldsymbol{\theta})}\bigr),
\)
where $f_{\boldsymbol{\theta}}$ is the probability density function (pdf) of an individual sample. 

The \emph{risk} incurred for estimating $\boldsymbol{\theta}$ as $\hat{\boldsymbol{\theta}}$ is given by
\begin{equation} \label{eq: risk}
R(\boldsymbol{\theta}, \hat{\boldsymbol{\theta}}) \coloneqq  \mathbb{E}_{\boldsymbol{y}_N^{(\boldsymbol{\theta})}} \bigl[L\bigl(\boldsymbol{\theta}, \hat{\boldsymbol{\theta}}(\boldsymbol{y}_N^{(\boldsymbol{\theta})})\bigr)\bigr],
\end{equation} 
where the expectation is with respect to the distribution $\mathbb{P}_{\boldsymbol{\theta}}$ of the observation vector $\boldsymbol{y}_N^{(\boldsymbol{\theta})}$.  The \emph{loss} function $L\colon \mathbb{R}^d \times \mathbb{R}^d \to \mathbb{R}_0^{+}$ is taken as
\begin{equation} \label{eq: loss}
    L\bigl(\boldsymbol{\theta},\hat{\boldsymbol{\theta}}(\boldsymbol{y}_N^{(\boldsymbol{\theta})})\bigr) = \bigl\lVert \boldsymbol{\theta} - \hat{\boldsymbol{\theta}}(\boldsymbol{y}_N^{(\boldsymbol{\theta})}) \bigr\rVert_2^2,
\end{equation}
%
which measures the squared error between the true parameter and its estimate.

The posterior distribution of $\ve{\theta}$ given the data is
\(
\mathbb{P}(\boldsymbol{\theta}\mid\boldsymbol{y}) = \frac{\mathbb{P}_{\boldsymbol{\theta}}(\boldsymbol{y}) \, \pi(\boldsymbol{\theta})}{\int_{\Theta} \mathbb{P}_{\boldsymbol{\theta}'}(\boldsymbol{y}) \, \pi(\boldsymbol{\theta}') \, d\boldsymbol{\theta}'}.
\)
The Bayes estimate minimizes the expected risk under the prior, and is given by
\begin{equation} \label{eq: bayes_estimator}
\begin{aligned}
\hat{\boldsymbol{\theta}}_{\mathrm{Bayes}}(\boldsymbol{y}) &= \arg\min_{\ve{\eta} \in \mathbb{R}^d} \mathbb{E}_{\boldsymbol{\theta} \sim \pi} \bigl[L(\boldsymbol{\theta}, \ve{\eta}) \mid \boldsymbol{y}\bigr] \\
&= \arg\min_{\ve{\eta} \in \mathbb{R}^d} \int_{\Theta} L(\boldsymbol{\theta}, \ve{\eta}) \, \mathbb{P}(\boldsymbol{\theta}\mid\boldsymbol{y}) \, d\boldsymbol{\theta}.
\end{aligned}
\end{equation}
In this paper, we focus on the Bayes TS estimator~\cite{garatti2008estimation,garatti2013new,BLCRCDC2022}, a simulation-driven approach that learns a functional mapping from observations to parameters, approximating the Bayes estimate~\eqref{eq: bayes_estimator}.

\subsection{TS Estimator and its Asymptotic Properties} \label{subsec: problem_statement}
In this section, we present the mathematical framework of the two-stage (TS) estimator, introduced in~\cite{garatti2008estimation,garatti2013new}, and then formalize the asymptotic properties used in our analysis by stating the definitions of consistency and asymptotic normality.

\begin{rem}
In the sequel, observations for $\tilde{\boldsymbol{\theta}}_i \in \Theta$ and $\boldsymbol{\theta} \in \Theta$ are denoted $\boldsymbol{y}_N^{(i)} =  (y_1^{(i)}, \ldots, y_N^{(i)})^{\top}$ and $\boldsymbol{y}_N^{(\boldsymbol{\theta})} =  (y_1^{(\boldsymbol{\theta})}, \ldots, y_N^{(\boldsymbol{\theta})})^{\top}$ respectively, and for the true system, $\boldsymbol{y}_N^{(0)} = (y_1^{(0)},\ldots,y_N^{(0)})^{\top}$.
\end{rem}

The TS estimator operates in two phases:
\begin{enumerate}
\item \textbf{Training Phase}: Sample parameters $\{\tilde{\boldsymbol{\theta}}_i\}_{i=1}^m$, $\tilde{\boldsymbol{\theta}}_i \stackrel{\mathrm{i.i.d.}}{\sim} \pi$. For each $\tilde{\boldsymbol{\theta}}_i$, $\mathcal{M}(\tilde{\boldsymbol{\theta}}_i)$ generates observations $\boldsymbol{y}_N^{(i)} = (y_1^{(i)}, \ldots, y_N^{(i)})^{\top}$, forming $\mathcal{D}_{\mathrm{tr}} = \{ (\boldsymbol{y}_N^{(i)}, \tilde{\boldsymbol{\theta}}_i) \}_{i=1}^m$. The TS estimator learns a mapping
\begin{equation} \label{eq: ts_estimator}
\hat{\boldsymbol{\theta}}_{\mathrm{TS}}(\cdot) = \arg\min_{\hat{\boldsymbol{\theta}}(\cdot) \in \Delta} \; \frac{1}{m} \sum_{i=1}^m L\bigl(\hat{\boldsymbol{\theta}}(\boldsymbol{y}_N^{(i)}), \tilde{\boldsymbol{\theta}}_i\bigr),
\end{equation}
where $\Delta = \{\hat{\boldsymbol{\theta}}\colon \mathbb{R}^N \to \mathbb{R}^d\}$. It has a composite structure:
\begin{equation} \label{eq: ts_composite}
\hat{\boldsymbol{\theta}}_{\mathrm{TS}} = g^* \circ h_N,
\end{equation}
%
with $\boldsymbol{h}_N\colon \mathbb{R}^N \to \mathbb{R}^n$, $n \ll N$, a compression function (e.g., sample quantiles), and $g^*\colon \mathbb{R}^n \to \mathbb{R}^d$ solving
\begin{equation} \label{eq:g_star}
  g^* = \arg\min_{g \in \mathcal{G}} \frac{1}{m} \sum_{i=1}^m L\bigl(g\bigl(\boldsymbol{h}_N(\boldsymbol{y}_N^{(i)})\bigr), \tilde{\boldsymbol{\theta}}_i\bigr),
\end{equation}
where $\mathcal{G}$ is a class of supervised learning models (e.g., kernel regressors).

\item \textbf{Implementation Phase}: Once the mapping $\hat{\boldsymbol{\theta}}_{\mathrm{TS}}$ has been trained, it is applied directly to new, real-world observations $\boldsymbol{y}_N^{(0)} = (y_1^{(0)}, \ldots, y_N^{(0)})^{\top}$ to produce an estimate of the true parameter:
\begin{equation} \label{eq: final_ts}
\hat{\boldsymbol{\theta}}_0 = \hat{\boldsymbol{\theta}}_{\mathrm{TS}}\bigl(\boldsymbol{y}_N^{(0)}\bigr).
\end{equation}\noeqref{eq: final_ts}
No optimization is performed at this stage; the mapping learned offline in the training phase is used to map the compressed observations to a parameter vector.
\end{enumerate}

{
\begin{rem}
Numerical comparisons of the TS estimator against standard parameter estimation approaches such as PEM, PF, KF and II can be found in~\cite{garatti2013new}, where the advantages of the TS approach compared to these approaches are demonstrated with several numerical examples, including a practical application.
\end{rem}}

The training phase of TS typically involves large values of $m$ and $N$, and in our asymptotic analysis we consider regimes in which $N$ (and subsequently $m$) grows.  In the limit, the estimator $\hat{\boldsymbol{\theta}}_{\mathrm{TS}}$ converges to a limit:
\begin{equation} \label{eq: limit_ts}
\bar{\boldsymbol{\theta}}_{\mathrm{TS}}(\cdot) = \bar{g} \circ \boldsymbol{h}_N(\cdot),
\end{equation} \noeqref{eq: limit_ts}
where
\begin{equation} \label{eq: limit_g_ts}
\bar{g} = \arg\min_{g \in \mathcal{G}} \lim_{m \to \infty} \lim_{N \to \infty} \frac{1}{m} \sum_{i=1}^m L\bigl(g\bigl(\boldsymbol{h}_N(\boldsymbol{y}_N^{(i)})\bigr), \tilde{\boldsymbol{\theta}}_i\bigr).
\end{equation} \noeqref{eq: limit_g_ts}
The limit $N \to \infty$ being taken before $m \to \infty$ reflects the assumption that the observation length grows faster than the number of parameter samples.  Consistency and normality will be proved for $\bar{\boldsymbol{\theta}}_{\mathrm{TS}}(\boldsymbol{y}_N^{(0)})$ as $N\to\infty$.

\textbf{Asymptotic Properties}: We establish consistency and asymptotic normality, accounting for three limits.

\begin{definition}[Consistency] \label{def: ts_consistency}
The TS estimator is said to be \emph{consistent} if, when observations $\boldsymbol{y}_N^{(0)}$ are generated from the true system $\mathcal{M}(\boldsymbol{\theta}_0)$,
\begin{equation} \label{eq: ts_consistency}
\bar{\boldsymbol{\theta}}_{\mathrm{TS}}(\boldsymbol{y}_N^{(0)}) \xrightarrow[\mathrm{a.s.}]{N \to \infty} \boldsymbol{\theta}_0, \quad \text{for all } \ve{\theta}_0 \in \Theta. \nonumber
\end{equation}
\end{definition}

\begin{definition}[Asymptotic Normality] \label{def: ts_normality}
The TS estimator is said to be \emph{asymptotically normal} if there exists a positive definite matrix $\ve{\Sigma} \in \mathbb{R}^{d \times d}$ such that, when observations $\boldsymbol{y}_N^{(0)}$ are generated from $\mathcal{M}(\boldsymbol{\theta}_0)$,
\begin{equation} \label{eq: ts_normality}
\sqrt{N} \left[ \bar{\boldsymbol{\theta}}_{\mathrm{TS}}(\boldsymbol{y}_N^{(0)}) - \boldsymbol{\theta}_0 \right] \xrightarrow[N \to \infty]{d} \mathcal{N}(\boldsymbol{0}, \ve{\Sigma}), \quad \text{for all } \ve{\theta}_0 \in \Theta. \nonumber
\end{equation}
\end{definition}

\begin{rem}
The need for three limits ($m \to \infty$ and $N \to \infty$ in training, and $N \to \infty$ in implementation) to define the asymptotic properties of TS distinguish it from standard point estimation, requiring a tailored analysis.
\end{rem}

In this paper, we seek to establish a simplified analysis of the TS estimator’s asymptotic properties, according to Definitions~\ref{def: ts_consistency} and~\ref{def: ts_normality}. In the following section, we shall state the choice of first and second stages of the TS estimator for which its asymptotic consistency and normality are established under simplified assumptions.
}

\section{Asymptotic Analysis of TS Estimator} \label{sec: Asymptotic analysis}
\vspace{-1 em}
In this section, we shall state the choice of the first and second stages of the TS estimator, followed by stating our main assumptions for its asymptotic analysis.

\textit{First stage of the TS estimator} \\
The first stage of the TS estimator is a data compression function, denoted by $\ve{h}_N\colon \mathbb{R}^N \to \mathbb{R}^n$, where $n \ll N$ is fixed. In this paper, we choose $\boldsymbol{h}_N$ as the function that computes sample quantiles as in Definition~\ref{def: sample quantile}. 

\begin{definition}[Sample quantiles] \label{def: sample quantile}
Let \(y_{(1)}^{(\boldsymbol{\theta})} \le \cdots \le y_{(N)}^{(\boldsymbol{\theta})}\) denote the order statistics of a sample \(y_1^{(\boldsymbol{\theta})},\ldots,y_N^{(\boldsymbol{\theta})}\).  Fix \(n \ll N\) and define levels \(\gamma_k = \frac{k}{n+1}\) for \(k=1,\dots,n\).  The empirical \(\gamma_k\)-quantile is \(y_{(\lfloor \gamma_k N\rfloor)}\), and the first-stage compression function collects these quantiles into
\begin{equation} \label{eq: first_stage_data}
\ve{h}_N(y_1^{(\boldsymbol{\theta})},\dots,y_N^{(\boldsymbol{\theta})}) = \bigl(y_{(\lfloor \gamma_1 N\rfloor)}^{(\boldsymbol{\theta})},\dots,y_{(\lfloor \gamma_n N\rfloor)}^{(\boldsymbol{\theta})}\bigr)^{\top}.
\end{equation}
\end{definition}


As $N \to \infty$, the sample quantiles typically converge to the $\gamma$-quantiles {(proved later, See Lemma~\ref{lemma: quantile convergence})}, which are defined as follows.

\begin{definition}[\(\gamma\)-quantiles]
For a parameter value \(\theta \in \Theta\) and a level \(\gamma \in [0,1]\), the \(\gamma\)-quantile of the distribution \(F_\theta\) is defined by
\[
  q^{(\theta)}(\gamma) := \inf\{x \in \mathbb{R}\colon F_\theta(x) \ge \gamma\}.
\]
\end{definition}

For the vector of quantile levels $\boldsymbol{\gamma}_n=(\gamma_{1},\dots,\gamma_{n})^{\top}$, define
\begin{equation} \label{eq: h_theta}
\ve{h}^{(\boldsymbol{\theta})}(\boldsymbol{\gamma}_n) = \bigl(q^{(\boldsymbol{\theta})}(\gamma_{1}),\dots,q^{(\boldsymbol{\theta})}(\gamma_{n})\bigr)^{\top}.
\end{equation}
The vector $\ve{h}^{(\boldsymbol{\theta})}(\boldsymbol{\gamma}_{n})$ collects the $\gamma$-quantiles corresponding to the order statistics extracted in the first stage.  

\textit{Second stage of the TS estimator} \\
%
%
The second stage of the TS estimator will be a supervised learning model that has a suitable functional representation $\ve{g}\colon \mathbb{R}^n \to \mathbb{R}^d$. In this paper, we choose $g$ to be a polynomial regressor, which is a functional of the form:
\begin{equation} \label{eq: supervised_representation}
\ve{g}(\ve{z}) = \boldsymbol{\beta}^{\top} \ve{J}(\ve{z}),
\end{equation}
where $\ve{J}\colon \mathbb{R}^{n} \to \mathbb{R}^{p}$ is a continuously differentiable feature map (e.g., monomials of fixed degree) and $\boldsymbol{\beta}$ is a $p \times d$ coefficient matrix.  In the scalar‑parameter case $d=1$, $\ve{\beta} \in \mathbb{R}^{p}$ and $g(z) = \boldsymbol{\beta}^{\top} \ve{J}(z)$ yields a scalar output.

Thus, $\ve{g}$ receives the data from the first stage, as given by~\eqref{eq: first_stage_data}, and then {outputs an estimate of $\boldsymbol{\theta}$ as} 
\begin{align} \label{eq: second_stage_TS}
\hat{\boldsymbol{\theta}}_{\text{TS}}((y_1^{(\boldsymbol{\theta})},\ldots,y_N^{(\boldsymbol{\theta})})^{\top})
&= \ve{g}(\ve{h}_N((y_1^{(\boldsymbol{\theta})},\ldots,y_N^{(\boldsymbol{\theta})})^{\top})) \nonumber \\
&= \boldsymbol{\beta}^{\top} \ve{J}(\ve{h}_N((y_1^{(\boldsymbol{\theta})},\ldots,y_N^{(\boldsymbol{\theta})})^{\top})).
\end{align}
Therefore, combining~\eqref{eq: first_stage_data} and~\eqref{eq: second_stage_TS}, the TS estimator is a functional of the form
\begin{equation} \label{eq: final_TS}
\hat{\boldsymbol{\theta}}_{\text{TS}} = \boldsymbol{\beta}^{\top} (\ve{J} \circ \ve{h}_N).  
\end{equation} \noeqref{eq: final_TS}

{
\begin{rem}[Choice of regression model]
Although we focus on polynomial feature maps in \eqref{eq: supervised_representation}, our consistency and normality results extend to any regression function $\ve{g}$ that is continuous and sufficiently expressive to approximate the mapping $\mathbf{z} \mapsto \ve{\theta}$ uniformly over $\Theta$.  For example, one may replace $\ve{J}$ by a neural network of enough capacity. The key requirement is that the bias error due to the function approximation vanishes faster than the variance error as $m, N \to \infty$. This is analogous to the universal approximation assumption used in nonparametric regression and prediction-error methods.
\end{rem}
}
\subsection{Main Assumptions} \label{subsec: assumptions} 
In this subsection, we state the main assumptions considered for the asymptotic analysis of the TS estimator. 

\begin{assumption} \label{assumption: autonomous}
The system \(\mathcal{M}(\boldsymbol{\theta})\) and true system are autonomous, \emph{i.e.}, do not have input signals.
\end{assumption}

{
\begin{assumption} \label{assumption: model}
For each \(\boldsymbol{\theta} \in \Theta\), \(\mathcal{M}(\boldsymbol{\theta})\) generates observations \(\boldsymbol{y}_N^{(\boldsymbol{\theta})} = (y_1^{(\boldsymbol{\theta})}, \ldots, y_N^{(\boldsymbol{\theta})})^{\top} \in \mathbb{R}^N\) on \((\Omega, \mathcal{F}, \mathbb{P})\), where, given \(\boldsymbol{\theta}\), \(y_j^{(\boldsymbol{\theta})} \stackrel{i.i.d.}{\sim} {f_{\boldsymbol{\theta}}}\) with conditional distribution \(\mathbb{P}_{\boldsymbol{\theta}}(\boldsymbol{y}_N^{(\boldsymbol{\theta})}) = \prod_{j=1}^N f_{\boldsymbol{\theta}}(y_j^{(\boldsymbol{\theta})})\), whose pdf \(f_{\boldsymbol{\theta}}\) and cumulative distribution function (cdf) \(F_{\boldsymbol{\theta}}\) are continuous.
\end{assumption}
}

\begin{assumption} \label{assumption: system}
The true system is \(\mathcal{M}(\boldsymbol{\theta}_0)\), with \(\boldsymbol{\theta}_0 \in \Theta\).
\end{assumption}




\begin{assumption} \label{assumption: identifiability}
There exists a \(\boldsymbol{\beta}^* \in \mathbb{R}^{p \times d} \) such that \(\boldsymbol{\beta}^{*\top} \ve{J}(\ve{h}^{(\boldsymbol{\theta})}(\boldsymbol{\gamma}_n)) = \boldsymbol{\theta}\) for each \(\boldsymbol{\theta} \in \Theta\), where $\ve{J}\colon \mathbb{R}^n \to \mathbb{R}^p$ is a continuously differentiable function, and \(\ve{h}^{(\boldsymbol{\theta})}(\boldsymbol{\gamma}_n) \in \mathbb{R}^n\) (cf. Eq.~\eqref{eq: h_theta}).
\end{assumption}

\begin{assumption} \label{assumption: invertibility}
\(\mathbb{E}_{\boldsymbol{\theta}}[\ve{J}(\ve{h}^{(\boldsymbol{\theta})}(\boldsymbol{\gamma}_n)) \ve{J}^{\top}(\ve{h}^{(\boldsymbol{\theta})}(\boldsymbol{\gamma}_n))]\) is invertible.
\end{assumption}

\begin{assumption} \label{assumption: F_diff}
\(F_{\boldsymbol{\theta}}(\cdot)\) has connected support, and \(F_{\boldsymbol{\theta}}^{-1}(\cdot)\) is differentiable for each \(\boldsymbol{\theta} \in \Theta\).
\end{assumption}

{
The following remarks on these assumptions are in order. 
\begin{enumerate}
    \item Autonomous models arise naturally when inputs are constant or unobserved--- as is typical in reliability or vibration problems.  Section~\ref{subsec: discussion non iid} shows that, appealing to standard prediction–error arguments, the theory extends to systems with external inputs.
    \item The i.i.d.~hypothesis is the classical baseline for large-sample analysis.  The same consistency and normality conclusions typically hold for weakly dependent, stationary data by replacing independence with common mixing~\cite{caines2018linear} or $m$-dependence~\cite{LJUNG1976121}. The continuity of $f_{\boldsymbol{\theta}}$ and $F_{\boldsymbol{\theta}}$ ensures that the quantiles are smooth functions of $\boldsymbol{\theta}$.
    \item Assuming that $\mathcal{M}(\theta_0)$ belongs to the model class defines a clear target parameter. 
    \item Assumption~\ref{assumption: identifiability} is the TS analogue of global identifiability, as considered in PEM: it requires that $\ve \theta$ be a linear functional of the feature vector.  The condition can be relaxed if $\ve{J}$ is replaced by a universal approximator with controlled error.
    \item Assumption~\ref{assumption: invertibility} prevents degeneracy in the training regression by ensuring the feature covariance matrix is nonsingular, mirroring the requirement of local identifiability in classical estimation.
    \item Assumption~\ref{assumption: F_diff}, on the connected support and a differentiable quantile map, is a standard regularity condition that justifies the Delta-method step; virtually all continuous distributions used in practice satisfy it.
\end{enumerate}
}





\subsection{Consistency of the TS Estimator}\label{sec: Consistency} 
\vspace{-0.5 em}
In this section, we derive the consistency and asymptotic normality of the TS approach. To derive the consistency, we first prove that the first stage, compression function $\ve{h}_N$, converges almost surely (a.s.) and consequently, also the second stage, which is a continuous map of the first stage,  converges a.s., thereby establishing the consistency of TS. The derivation of asymptotic normality follows similar lines. 



\textbf{Note:} For ease of exposition we present the proofs for the \emph{scalar} parameter case (\(d = 1\)).  In this setting the regression coefficient \(\boldsymbol{\beta}\) (and its finite-sample estimate \(\hat{\boldsymbol{\beta}}\)) is a vector in \(\mathbb{R}^{p}\).  The results easily extend to the multivariate case \(\boldsymbol{\theta} \in \mathbb{R}^{d}\) by treating \(\boldsymbol{\beta}\) as a \(p \times d\) matrix and applying the analysis component-wise.




\subsubsection{Almost sure convergence of empirical quantiles}\label{subsec: Quantile a.s. convergence}

First, we prove that the order statistics  converge a.s. to $\gamma$-quantiles. The key result is the following:

\begin{lemma}\label{lemma: quantile convergence}
Under Assumptions~\ref{assumption: autonomous}-\ref{assumption: model}, for each $\gamma \in [0,1]$ and $\theta \in \Theta$, $y_{(\lfloor \gamma N \rfloor )}^{(\theta)} \stackrel{a.s.}{\rightarrow} q^{(\theta)}(\gamma)$ as $N \rightarrow \infty$.
\end{lemma}
\begin{pf}
See Appendix~\ref{appendix: proof_lemma_1}. \hfill $\blacksquare$
\end{pf}

Based on this lemma, $y_{(\lfloor \gamma N \rfloor)}^{(\theta)} \stackrel{a.s.}{\to} q^{(\theta)}(\gamma)$ as $N \to \infty$ for each fixed $\gamma \in [0,1]$. Then, it immediately follows that the first stage of TS converges a.s. to the true quantiles.

\subsubsection{Strong consistency of the TS estimator}\label{subsec: strong consistency of the TS}

We study the TS estimator on a fresh test sample $\boldsymbol{y}_N^{(0)}=(y_1^{(0)},\ldots,y_N^{(0)})^{\top}$,
$y_j^{(0)}\stackrel{\mathrm{i.i.d.}}{\sim}\mathbb{P}_{\theta_0}$,  generated by the true system $\mathcal{M}(\theta_0)$.
The finite-sample estimate is $\hat{\theta}_0 =\hat{\boldsymbol{\beta}}^{\top}\ve{J}\bigl(\ve{h}_N(\boldsymbol{y}_N^{(0)})\bigr)$, with $\hat{\boldsymbol{\beta}}$ obtained by solving
\begin{equation} \label{eq: TS_opt}
    \hat{\boldsymbol{\beta}} = \amin{\boldsymbol{\beta} \in \mathbb{R}^p } \frac{1}{m} \sum_{i=1}^{m} (\tilde{\theta}_i - \boldsymbol{\beta}^{\top} \ve{J}(\ve{h}_N(\boldsymbol{y}_N^{(i)})))^2.
\end{equation}
The solution to~\eqref{eq: TS_opt} is given by
\begin{equation}\label{eq: beta_N_M}
\begin{split}
        \hat{\boldsymbol{\beta}} &= \bigg[\sum_{i=1}^{m} \ve{J}(\ve{h}_N(\boldsymbol{y}_N^{(i)})) \ve{J}^{\top}(\ve{h}_N(\boldsymbol{y}_N^{(i)}))\bigg]^{-1} \\  &\quad \quad \cdot \sum_{i=1}^{m} \ve{J}(\ve{h}_N(\boldsymbol{y}_N^{(i)}))\tilde{\theta}_i.
\end{split}
\end{equation}
The limiting estimate \(\bar{\boldsymbol{\theta}}_{\text{TS}}(\boldsymbol{y}_N^{(0)})\) (cf. Eq.~\eqref{eq: limit_ts} and~\eqref{eq: limit_g_ts}) is given by 
\[
  \bar{\theta}_0
  =\bar{\boldsymbol{\beta}}^{\top}\ve{J}\!\bigl(\ve{h}_N(\boldsymbol{y}_N^{(0)})\bigr), \, \text{ where }
  \bar{\boldsymbol{\beta}}
  =\lim_{m\to\infty}\lim_{N\to\infty}\hat{\boldsymbol{\beta}}.
\]

Our aim is to prove:
\begin{align} 
&\text{(i) } \bar{\theta}_0 \xrightarrow[N\to\infty]{\mathrm{a.s.}} \theta_0. \label{eq: aim_to_prove_const} \\
 &\text{(ii) }  \sqrt{N}\bigl(\bar{\theta}_0-\theta_0\bigr)
  \xrightarrow[N\to\infty]{d}\mathcal{N}(0,\Sigma_{\mathrm{TS}}).  \label{eq: aim_to_prove_normality}
  \end{align}
The next lemma provides the almost-sure convergence of the first-stage (compression) and serves as the starting point of the argument.

\begin{lemma} \label{lemma: a.s. converegence continuous function}
Let $\ve{J}\colon \mathbb{R}^n \to \mathbb{R}^p $ be a continuous function and $\ve{x}_1, \ldots, \ve{x}_N$ be i.i.d.~random variables defined on a probability space $(\Omega,\mathcal{A},\mathbb{P})$ with $\ve{x}_i\colon \Omega \to \mathbb{R}^n$. Then, if $\ve{x}_N \stackrel{a.s.}{\to} \ve{x}$, it holds that $\ve{J}(\ve{x}_N) \stackrel{a.s.}{\to} \ve{J}(\ve{x})$.
\end{lemma}
\begin{pf}
\vspace{-1.5 em}
See~\cite[page~8]{van2000asymptotic}. $\quad \quad \quad \quad \quad \quad \quad \quad \quad \quad \quad \quad \blacksquare$
\end{pf}

As the result of Lemma~\ref{lemma: a.s. converegence continuous function}, 
\begin{equation} \label{eq: J_convergence}
 \ve{J}(\ve{h}_N(\boldsymbol{y}_N^{(\theta)})) \stackrel{a.s.}{\to} \ve{J}(\ve{h}^{(\theta)}(\boldsymbol{\gamma}_n)) \text{ for each } \theta \in \Theta.    
\end{equation}

To establish the strong consistency of the TS estimator (c.f.~Eq.~\eqref{eq: aim_to_prove_const}), we begin by showing in Theorem~\ref{thm: Convergence of TS} together with Eq.~\eqref{eq: J_convergence} that, as $N \to \infty$ and $m \to \infty$, we have that $\hat{\boldsymbol{\beta}} \to \bar{\boldsymbol{\beta}} \in \mathbb{R}^p $ with probability $1$, where 
\begin{equation}\label{eq: bar_beta}
\begin{split}
    &\bar{\boldsymbol{\beta}} = \mathbb{E}_{\theta}[\ve{J}(\ve{h}^{(\theta)}(\boldsymbol{\gamma}_n)) \ve{J}^{\top}(
 \ve{h}^{(\theta)}(\boldsymbol{\gamma}_n))]^{-1}
    \mathbb{E}_{\theta}[\theta \ve{J}(\ve{h}^{(\theta)}(\boldsymbol{\gamma}_n))].
\end{split}
\end{equation}

\begin{theorem}\label{thm: Convergence of TS}
Under Assumptions~\ref{assumption: autonomous}-\ref{assumption: model} and Lemma~\ref{lemma: a.s. converegence continuous function}, $\lim_{m \to \infty} \lim_{N \to \infty} \hat{\boldsymbol{\beta}} \stackrel{a.s.}{\to} \bar{\boldsymbol{\beta}}$. 
\end{theorem}
\begin{pf}
See Appendix~\ref{appendix: proof_thm_1}. \hfill $\blacksquare$
\end{pf}

Finally, applying Theorem~\ref{thm: Convergence of TS}, the strong consistency of the TS estimator is established in Theorem~\ref{thm: strong consistency of TS}. 

\begin{theorem}[Strong consistency]\label{thm: strong consistency of TS}
Under Assumptions~\ref{assumption: autonomous}-\ref{assumption: invertibility}, $\bar{\theta}_{0} = \bar{\theta}_{\text{TS}}(\boldsymbol{y}^{(0)}_{N}) = 
\bar{\boldsymbol{\beta}}^{\top} \ve{J}(\ve{h}_N(\boldsymbol{y}_{N}^{(0)})) \stackrel{a.s.}{\to} \theta_0$ as $N \to \infty$.
\end{theorem}
\vspace{-1 em}
\begin{pf}
    See Appendix~\ref{appendix: proof_thm2}. \hfill $\blacksquare$
\end{pf}

Now, we shall establish the asymptotic normality of the TS estimator. 




\subsection{Asymptotic Normality of the TS estimator} \label{subsec: Asymptotic normality}
\vspace{-1 em}

Having established the consistency  of the TS estimator, we now establish its asymptotic normality, in the form of Eq.~\eqref{eq: aim_to_prove_normality}.  The argument proceeds in three steps:

\begin{enumerate}
\item derive a central-limit theorem (CLT) for the empirical quantiles that form the first stage of TS;
\item apply the Delta method (Lemma~\ref{lemma: Delta method}) together with Lemma~\ref{Lemma: quantile converge to inv cdf} to transfer the CLT to the vector $\ve{h}_N(\boldsymbol{y}_N^{(\theta)})$;
\item propagate this limit through the continuous second stage to obtain the asymptotic distribution of the TS estimate.
\end{enumerate}

Throughout this subsection we keep the notation introduced in Section~\ref{subsec: Quantile a.s. convergence}.  For each $x\in\mathbb{R}$ define the empirical cdf
\begin{equation}\label{eq: F_N}
F_N^{(\theta)}(x) = \frac{1}{N} \sum_{i=1}^N \mathds{1}\{y_i^{(\theta)} \le x\}
            = \frac{1}{N} \sum_{i=1}^N Z_i^{(\theta)},
\end{equation}
where the $Z_i^{(\theta)}$'s are i.i.d.~Bernoulli random variables with mean $F_{\theta}(x)$ and variance $F_{\theta}(x) (1-F_{\theta}(x))$. Hence, we can apply the central limit theorem to~\eqref{eq: F_N} to obtain
\begin{equation}\label{eq: Ascov F_N}
\sqrt{N} (F_N^{(\theta)}(x) - F_{\theta}(x)) \stackrel{d}{\to} \mathcal{N}(0,F_{\theta}(x)(1-F_{\theta}(x))). 
\end{equation}
Since we want to evaluate the asymptotic distribution of the sample quantiles, we need to analyze the asymptotic distribution of $F_{\theta}^{-1}(F_N^{(\theta)}(x))$, as seen later. Under Assumption~\ref{assumption: F_diff}, the Delta method~\cite{ferguson2017course,van2000asymptotic} can be applied to~(\ref{eq: Ascov F_N}) to obtain the asymptotic distribution of $F^{-1}_{\theta}(F_N^{(\theta)}(x))$.
Once we have this asymptotic distribution, we can easily extend it to obtain the asymptotic distribution of $y_{(\lfloor \gamma N  \rfloor)}^{(\theta)}$. To this end, we will use the following version of the Delta method. 

\begin{lemma}[Delta method]\label{lemma: Delta method}
 Let $g\colon \mathbb{R} \to \mathbb{R}$ is a differentiable function. Let $x_N$, $N \in \mathbb{N}$ be i.i.d.~random variables defined on a probability space $(\Omega, \mathcal{A}, \mathbb{P})$ such that $N^b (x_N - a) \stackrel{d}{\to} x$ for some constants $b>0$ and $a$. Then,
    \begin{equation}
        N^b (g(x_N) - g(x)) \stackrel{d}{\to} g'(a)(x).
    \end{equation}
\end{lemma}
\vspace{-2 em}
\begin{pf}
    See~\cite[page~26, Theorem~3.1]{van2000asymptotic}. \hfill $\blacksquare$
\end{pf}
\vspace{-1em}
As we can notice from Lemma~\ref{lemma: Delta method}, we require to compute the derivative of $F^{-1}_{\theta}$ to derive the asymptotic distribution of $F^{-1}_{\theta}(F_N^{(\theta)}(x))$. We can compute this derivative as follows:
Let $t = F^{-1}_{\theta}(z)$, and let $f_{\theta}$ be the corresponding pdf. Then, $z = F_{\theta}(t)$ implies that $dz = f_{\theta}(t) dt$, hence
$\frac{dt}{dz} = \frac{1}{f_{\theta}(t)} = \frac{1}{f_{\theta}(F^{-1}_{\theta}(z))}$.
Therefore, 
\begin{equation} \label{eq: derivative of F theta inv}
    (F^{-1}_{\theta})'(z) = \frac{1}{f_{\theta}(F^{-1}_{\theta}(z))}.
\end{equation}
Using~\eqref{eq: derivative of F theta inv} together with $b=1/2$ in Lemma~\ref{lemma: Delta method} and~\eqref{eq: Ascov F_N} yields
\begin{multline}\label{eq: Ascov of inv cdf}
\sqrt{N} (F^{-1}_{\theta}(F_N^{(\theta)}(x)) - F^{-1}_{\theta}(F_{\theta}(x)))\\
\stackrel{d}{\to} \mathcal{N}\left(0,\frac{F_{\theta}(x)(1-F_{\theta}(x))}{f_{\theta}^2(x)}\right).
\end{multline}
In fact, the following result also holds:
\begin{lemma}\label{Lemma: quantile converge to inv cdf}
    $y_{(\lfloor \gamma N \rfloor)}^{(\theta)}  - F^{-1}_{\theta}(F_N^{(\theta)}(x))\stackrel{a.s.}{\to} 0$ for $\gamma = F_{\theta}(x)$. 
\end{lemma}
\vspace{-1 em}
\begin{pf}
    The proof is similar to that of Lemma~\ref{lemma: quantile convergence}. $\quad \blacksquare$
\end{pf}

Lemma~\ref{Lemma: quantile converge to inv cdf} implies that for each $x \in \mathbb{R}$, there is a $\gamma = F_{\theta}(x)$ such that $y_{(\lfloor \gamma N \rfloor)}^{(\theta)} - F^{-1}_{\theta}(F_N^{(\theta)}(x)) \stackrel{a.s.}{\to} 0$. Furthermore, $y_{(\lfloor \gamma N \rfloor)}^{(\theta)} -F^{-1}_{\theta}(F_N^{(\theta)}(F^{-1}_{\theta}(\gamma))) \stackrel{a.s.}{\to} 0$ for each $\gamma \in [0,1]$.

With this observation and upon applying Eq.~\eqref{eq: Ascov of inv cdf} with Slutsky's theorem~\cite{ferguson2017course}, we finally obtain
\begin{equation}\label{eq: Ascov of quantiles}
    \sqrt{N}(y_{(\lfloor \gamma N \rfloor)}^{(\theta)}- F^{-1}_{\theta}(\gamma)) \stackrel{d}{\to} \mathcal{N}\left(0,\frac{\gamma (1-\gamma)}{(f_{\theta}(F^{-1}_{\theta}(\gamma)))^2}\right).
\end{equation}

Since $y_1^{(\theta)},\ldots,y_N^{(\theta)}$ are i.i.d.~with distribution $F_{\theta}$, Eq.~\eqref{eq: Ascov of quantiles} implies that
\begin{equation} \label{eq: Asdist of quantiles}
    \sqrt{N}(\ve{h}_N(\boldsymbol{y}_{N}^{(\theta)}) - \ve{h}^{(\theta)}(\boldsymbol{\gamma}_n)) \stackrel{d}{\to} \mathcal{N}(\ve{0}, \ve{\Sigma}),
\end{equation}
where $\boldsymbol{\gamma}_n=(\gamma_1,\ldots,\gamma_n)^{\top}$, and
\begin{equation} \label{eq:Ascov matrix quantiles}
    \ve{\Sigma} = \text{diag}\left( \frac{\gamma_1 (1-\gamma_1)}{(f_{\theta}(F_{\theta}^{-1}(\gamma_1)))^2},\ldots, \frac{\gamma_n (1-\gamma_n)}{(f_{\theta}(F_{\theta}^{-1}(\gamma_n)))^2} \right).
\end{equation}

Therefore, the output of the compression function $\ve{h}_N$, the first stage of TS, is asymptotically normal. Now, it remains to show that the TS estimator is also asymptotically normal. This immediately follows from the definition of TS followed by the Delta method (see Lemma~\ref{lemma: Delta method}), as established in the following theorem.  

\begin{theorem}[Asymptotic Normality] \label{thm: Asymptotic normality}
    Under Assumptions~\ref{assumption: autonomous}-\ref{assumption: F_diff},
    $\sqrt{N}(\hat{\theta}_{0}-\theta_0) \stackrel{d}{\to} \mathcal{N}(0,\Sigma_{\text{TS}})$, where $\Sigma_{\text{TS}} = \bar{\boldsymbol{\beta}}^{\top} \nabla \ve{J}(\ve{h}^{(0)}) \ve{\Sigma} \nabla \ve{J}(\ve{h}^{(0)})^{\top} \bar{\boldsymbol{\beta}}$, and $\nabla \ve{J}(\ve{h}^{(0)}) \in \mathbb{R}^{m \times n}$ is the Jacobian matrix of $\ve{J}$ evaluated at $\ve{h}^{(0)}(\boldsymbol{\gamma}_n)$.
\end{theorem}
\begin{pf}
See Appendix~\ref{appendix: proof_thm_3}. 
\end{pf}

\begin{rem} \label{rem: Asymptotic normality}
The asymptotic normality has been established in Theorem~\ref{thm: Asymptotic normality} for the case when $\theta$ is a scalar. However, the analysis can be easily extended to the multivariate case by noting that $\bar{\boldsymbol{\beta}} \in \mathbb{R}^{p \times d}$ if $\ve{\theta} \in \mathbb{R}^{d}$. In this case, $\ve{\Sigma}_{\text{TS}}$ is a $d \times d$ matrix. 
\end{rem}

{
\subsection{Discussion on the Non i.i.d. Case} \label{subsec: discussion non iid}

In this subsection we briefly outline how the analysis of the TS estimator can be extended when the data–generating mechanism is non i.i.d. Here, Assumptions~\ref{assumption: autonomous}–\ref{assumption: model} are relaxed: the model structure $\mathcal{M}(\theta)$ and the true system $\mathcal{M}(\theta_0)$ have external inputs and generate stationary, dependent data.  A convenient first stage is given by the least-squares coefficients of an $\operatorname{ARX}(n_a,n_b)$ model fitted to each input–output record.

Let \(n = n_a + n_b\) be the total number of ARX coefficients.  
For each training pair, collect the input–output record in
\(\boldsymbol{z}_N^{(i)} = (u_1^{(i)},y_1^{(i)},\ldots,u_N^{(i)},y_N^{(i)})^\top\) and define
\vspace{-1 em}
\begin{equation}
\begin{split}
  h_N(\boldsymbol{z}_N^{(i)})
  &= \arg\min_{\boldsymbol{\xi}^{(i)}\in\mathbb{R}^{n}}
    \sum_{t=n_a+1}^{N}
      \bigl(y_t^{(i)} - \boldsymbol{\phi}_t^{\top}\boldsymbol{\xi}^{(i)}\bigr)^2,\\
  \boldsymbol{\phi}_t
  &= (y_{t-1}^{(i)},\ldots,y_{t-n_a}^{(i)},u_{t-1}^{(i)},\ldots,u_{t-n_b}^{(i)})^\top.
  \end{split}
\end{equation}
Standard prediction-error theory~\cite{Ljung:99}, together with a persistently exciting input of order~\(n\) for every \(\tilde{\boldsymbol{\theta}}_i\in\Theta\), implies
\[
  h_N(\boldsymbol{z}_N^{(i)}) \;\xrightarrow[N\to\infty]{\mathrm{a.s.}}\; (\boldsymbol{\xi}^{(i)})^\ast .
\]
Combining this limit with Assumption~\ref{assumption: identifiability} and the continuity of the second-stage map $\ve{g}$, the proof of strong consistency in Section~\ref{sec: Consistency} carries over verbatim, with Lemma~\ref{lemma: quantile convergence} replaced by the ARX limit above.  Lemmas 2–3 and Theorems 2–3 in this paper are still required for the remainder of the argument.

Analogously, prediction-error results give the central limit theorem
\[
  \sqrt{N}\bigl(\ve{h}_N(\boldsymbol{z}^{(0)}) - (\boldsymbol{\xi}^{(0)})^\ast\bigr)
  \;\xrightarrow{d}\;
  \mathcal{N}\!\bigl(\boldsymbol{0},\,
    \sigma^2\bigl(\mathbb{E}[\boldsymbol{\phi}_t\boldsymbol{\phi}_t^{\top}]\bigr)^{-1}\bigr),
\]
which, via the Delta method, yields the asymptotic normality of the TS estimator under non-i.i.d. assumptions.

\begin{rem}
Alternatively, we can replace independence with weak-dependence assumptions (mixing/$m$-dependence) so that the asymptotic theory for order statistics and our quantile-based first stage still apply~\cite{caines2018linear,mohri2008rademacher,vidyasagar2013learning}.
\end{rem}
}

{
\section{Numerical Validation of Asymptotic Results} \label{sec: numerical validation}

In this section, we corroborate the asymptotic results of Section~3 with a simple numerical illustration. Specifically, we consider the problem of estimating the signal-to-noise ration (SNR) of a system \(\mathcal{M}({\theta})\), where \( \theta = \frac{\mu^2}{\sigma^2}\) is the SNR, with known mean \(\mu = 5\) and true parameter \(\theta_0 = 5\), so \(\sigma_0^2 = \frac{\mu^2}{\theta_0} = 5\). The observations \(\boldsymbol{y}_N^{(0)} = (y_1^{(0)}, \ldots, y_N^{(0)})^{\top}\) are generated according to
\begin{equation} \label{eq: snr_model}
y_t = \mu + e_t, \quad e_t \stackrel{i.i.d.}{\sim} \mathcal{N}(0, \sigma^2),
\end{equation}
implying that \(y_t \stackrel{i.i.d.}{\sim} \mathcal{N}(\mu, \sigma^2)\), with a uniform prior \(\pi(\theta) \sim \mathcal{U}[2, 10]\). The TS estimator uses sample quantiles with \(n=5\) as the compression function \(h_N\), selecting order statistics at indices \(s = \lfloor j \cdot \frac{N}{n+1} \rfloor\), \(j = 1, \ldots, n\). Training involves \(m \in \{100, 1000, 10000\}\) parameter samples and observation lengths \(N \in \{100, 1000, 10000\}\). The mapping \(g\) is learned via polynomial regression (degree $=2$). In the implementation phase, the trained TS estimator is evaluated on test sequences whose length matches the number of training samples \(N\). The  mean squared error (MSE), given by, \(\frac{1}{R}\sum_{r=1}^{R}\bigl(\hat{\theta}^{(r)}-\theta_0\bigr)^2\) is computed over \(R= 100\) Monte-Carlo runs.

The Fisher Information for \(\sigma^2\) is
\[
I(\sigma^2) = \mathbb{E}\left[ \left( \frac{\partial \log \mathbb{P}_{\boldsymbol{\theta}}(\boldsymbol{y}_N^{(\boldsymbol{\theta})})}{\partial \sigma^2} \right)^2 \right] = \frac{N}{4 (\sigma^2)^2}.\]
Transforming this result to \(\theta\), with \(\frac{\partial \sigma^2}{\partial \theta} = -\frac{\mu^2}{\theta^2}\), yields the Fisher Information for $\theta$ as
\begin{equation} \label{eq: snr_fisher_theta}
I(\theta) = I(\sigma^2) \cdot \left( \frac{\partial \sigma^2}{\partial \theta} \right)^{-2} = \frac{N}{4 (\sigma^2)^2} \cdot \frac{\theta^4}{\mu^4} = \frac{N \theta^2}{4 \mu^4}.
\end{equation}
Thus, the Cramer-Rao Bound (CRB) for $\theta$ is given by \(\text{CRB}(\theta) = \frac{1}{F(\theta)} = \frac{4 \mu^4}{N \theta^2} = \frac{4 \theta \mu^2}{N}.\)
For \(\theta_0 = 5\) and \(\mu = 5\), \(\text{CRB}(\theta) = \frac{4 \cdot 5 \cdot 5^2}{N} = \frac{500}{N}\). Figures~\ref{fig: asym_valid_consistency} and \ref{fig: asym_valid_normality} corroborate, via Monte Carlo simulations of this system, the consistency and asymptotic normality of the TS estimator respectively.  
\begin{figure*}[h]
\begin{minipage}{0.45 \textwidth}
\centering
\includegraphics[width=0.7\columnwidth]{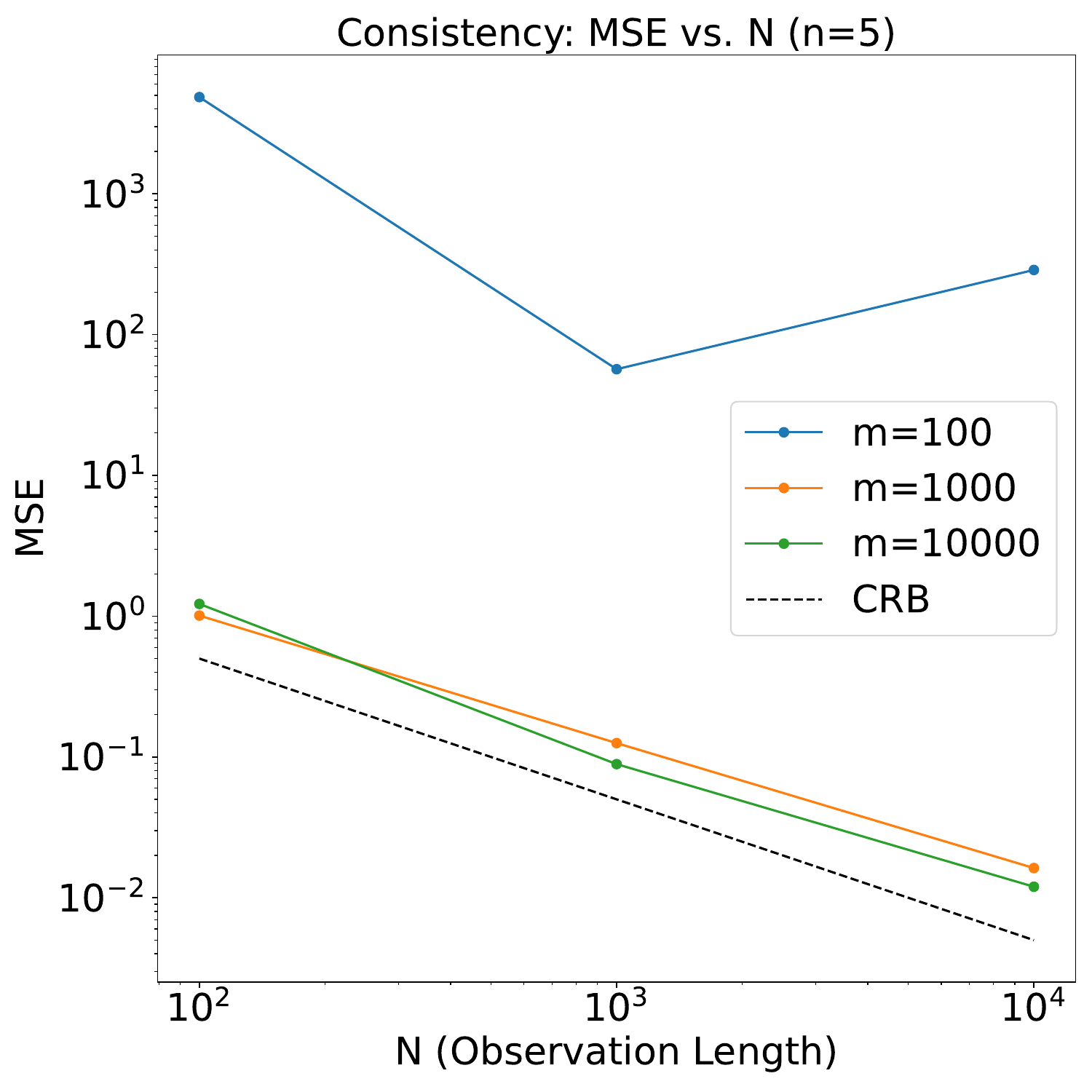}
\caption{Consistency of TS across training sizes \(m\) and lengths \(N\). Each marker aggregates 100 runs. MSE decreases with larger \((m,N)\) and approaches the CRB (\(500/N\), shown for reference), indicating near-efficiency in large samples.}
\label{fig: asym_valid_consistency}
\end{minipage}
\hfill
\begin{minipage}{0.45 \textwidth}
\centering
\includegraphics[width=0.7\columnwidth]{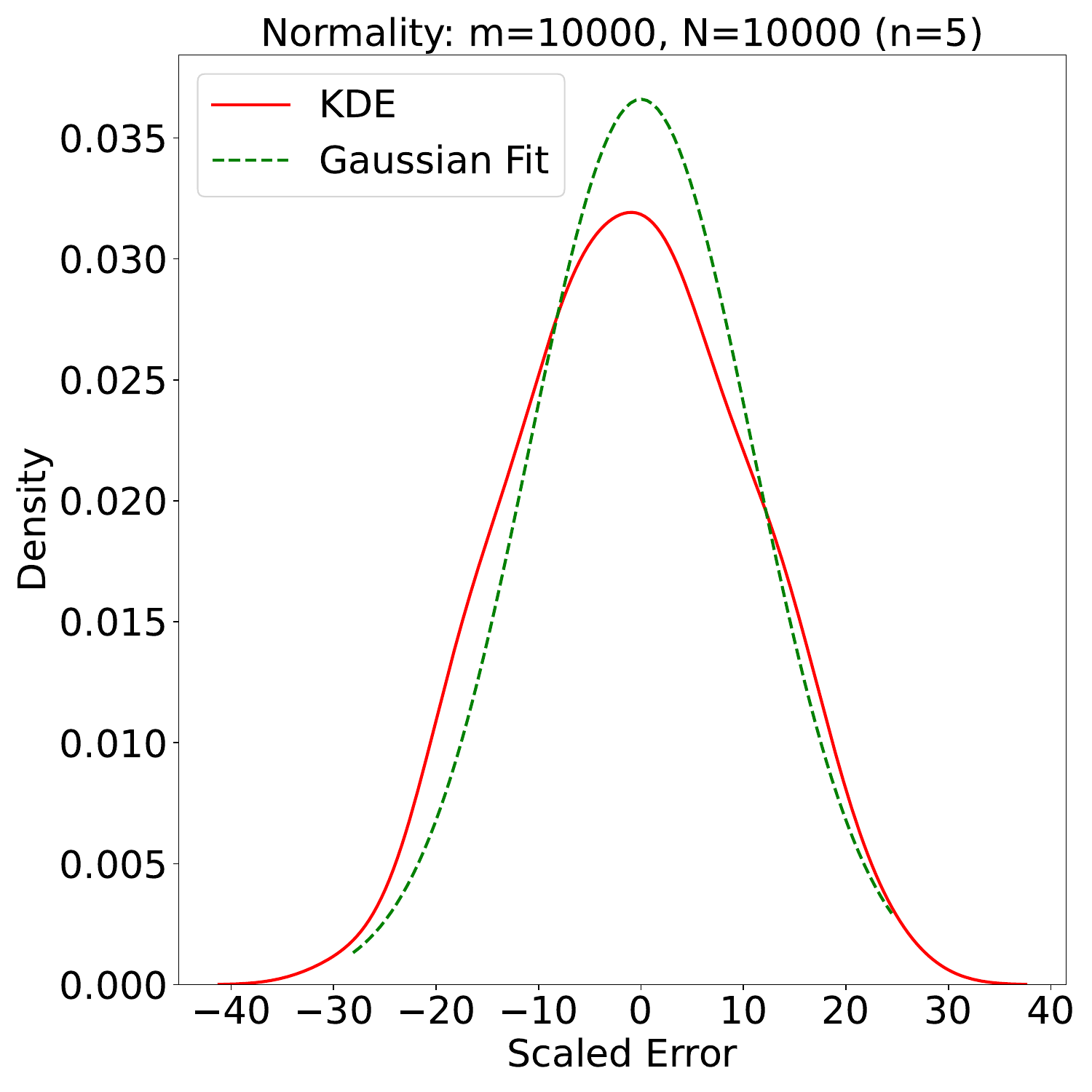}
\caption{Asymptotic normality: empirical distribution of \(\sqrt{N}\,(\hat{\boldsymbol{\theta}}_N-\theta_0)\) for large \(m,N\). The standardized errors concentrate and align with a Gaussian law, consistent with Section~3.3.}
\label{fig: asym_valid_normality}
\end{minipage} 
\end{figure*}


The consistency plot validates our result on the consistency of TS under i.i.d.~assumptions. As \( m \) and \( N \) increase to \( 10^4 \), the MSE decreases, approaching the CRB, which indicate near-efficiency and consistency as \( N \to \infty \), \( m \to \infty \). The normality plot confirms that for large \( m \) and \( N \), the distribution of \(\sqrt{N} (\hat{\boldsymbol{\theta}}_N - \theta_0)\) is approximately Gaussian, corroborating our theoretical results from Section 3.3.

}

\section{Conclusions} \label{sec: Conclusion}

{In this paper, we have established that the TS estimator is \emph{consistent} and \emph{asymptotically normal} in the i.i.d.\ setting with quantile-based compression and a linear-in-features second stage, and we have derived a closed-form expression for its asymptotic covariance. We have also briefly outlined how our analysis can be extended to stationary, weakly dependent data.

For future work, we aim to provide a rigorous treatment of the non-i.i.d.\ case, robustness under model misspecification, non-asymptotic (finite-sample) guarantees, and principled design of the compression stage for statistical efficiency. }

\appendix
\section{Appendix: Proofs of Technical Results}
The proofs of the lemmas and theorems stated in the main text are collected in this section of the appendix. 


\subsection{Proof of Lemma~\ref{lemma: quantile convergence}} \label{appendix: proof_lemma_1}
\text{For each } $\varepsilon >0$, let $A_{k,\varepsilon}~\coloneqq \{\vert y_{(\lfloor \gamma N \rfloor)}^{(\theta)} - q^{(\theta)}(\gamma)\vert \leq \varepsilon, \, \text{ for every } N\geq k  \}$ be an event in $\mathcal{A}$. To show that $y_{(\lfloor \gamma N \rfloor )}^{(\theta)} \stackrel{a.s}{\rightarrow} q^{(\theta)}(\gamma)$ as $N \rightarrow \infty$, we need to establish that for every $\varepsilon >0$, $\mathbb{P}(A_{k,\varepsilon}) \rightarrow 1$ as $k \to \infty$.
Notice that
\medskip
  \[        \mathbb{P}(A_{k,\varepsilon}^{c}) = \mathbb{P}(\vert{y_{(\lfloor \gamma N \rfloor)}^{(\theta)} - q^{(\theta)}(\gamma)\vert} > \varepsilon, 
      \text{ for some } N\geq k ).\]
\vspace{-1 em}
Then,
\begin{equation}\label{eq: Am_c}
\vspace{-1.4 em}
\begin{split}
\mathbb{P}(A_{k,\varepsilon}^{c}) &= \mathbb{P}(\cup_{N=k}^{\infty} \{\vert{y_{(\lfloor \gamma N \rfloor)}^{(\theta)} - q^{(\theta)}(\gamma)\vert} > \varepsilon\})  \\
&\stackrel{(a)}{\leq} \sum_{N=k}^{\infty} \mathbb{P}(\vert{y_{(\lfloor \gamma N \rfloor)}^{(\theta)} - q^{(\theta)}(\gamma)\vert} > \varepsilon) \\
&\stackrel{(b)}{=} \sum_{N=k}^{\infty}\mathbb{P}({y_{(\lfloor \gamma N \rfloor)}^{(\theta)}- q^{(\theta)}(\gamma)} > \varepsilon) \ + \\
&\quad \quad \quad \quad  \mathbb{P}({y_{(\lfloor \gamma N \rfloor)}^{(\theta)}- q^{(\theta)}(\gamma)} < - \varepsilon),
    \end{split}
\end{equation}
\medskip 
where $(a)$ follows from the union bound and $(b)$ follows from the fact that the events $\{y_{(\lfloor \gamma N \rfloor)}^{(\theta)} - q^{(\theta)}(\gamma) > \varepsilon\}$ and $\{y_{(\lfloor \gamma N \rfloor)}^{(\theta)}- q^{(\theta)}(\gamma) < - \varepsilon\}$ are disjoint. 

Now, let us evaluate $\mathbb{P}({y_{(\lfloor \gamma N \rfloor)}^{(\theta)} - q^{(\theta)}(\gamma)} > \varepsilon)$. 
To this end, observe that
\vspace{-2 em}
\begin{equation}\label{eq: observe_quantile_charac}
\begin{split}
     y_{(l)}^{(\theta)} &\leq x \text{ if and only if } \sum_{j=1}^{N} \mathds{1}\{y_j^{(\theta)} \leq x\} \geq l,\\
      y_{(l)}^{(\theta)} &\geq x \text{ if and only if } \sum_{j=1}^{N} \mathds{1}\{y_j^{(\theta)} \geq x\} \geq N-l+1,
      \end{split}
  \end{equation}
Then $\mathbb{P}({y_{(\lfloor \gamma N \rfloor)}^{(\theta)} - q^{(\theta)}(\gamma)} > \varepsilon)$ can be bounded as
{
    \begin{equation*}
    \begin{split}
\, &\mathbb{P}({y_{(\lfloor \gamma N \rfloor)}^{(\theta)} - q^{(\theta)}(\gamma)} > \varepsilon) \\
&\qquad \stackrel{(c)}{=} \mathbb{P}\bigg( \frac{1}{N} \sum_{j=1}^N \mathds{1}\{y_j^{(\theta)} \geq q^{(\theta)}(\gamma) +\varepsilon\} \geq \\
&\qquad \qquad \qquad \qquad 1- \frac{\lfloor \gamma N \rfloor}{N}+ \frac{1}{N}\bigg) \\
&\qquad \leq \mathbb{P}\bigg( \frac{1}{N} \sum_{j=1}^N \mathds{1}\{y_j^{(\theta)} \geq q^{(\theta)}(\gamma) +\varepsilon\} \geq 1- \frac{\lfloor \gamma N \rfloor}{N}\bigg) \\
&\qquad =\mathbb{P}\bigg(\frac{1}{N} \sum_{j=1}^N (\mathds{1}\{y_j^{(\theta)} \geq q^{(\theta)}(\gamma) +\varepsilon\} \\
 &\quad \quad \quad \quad \quad - \mathbb{E}[\mathds{1}\{y_j^{(\theta)} \geq q^{(\theta)}(\gamma) +\varepsilon\}]) \\
&\qquad \quad \quad \quad \geq 1-\frac{\lfloor \gamma N \rfloor}{N} -
\mathbb{E}[\mathds{1}\{y_j^{(\theta)} \geq q^{(\theta)}(\gamma) +\varepsilon\}]\bigg) \\
&\qquad \stackrel{(d)}{\leq} \exp\bigg(-2 N \bigg(F(q^{(\theta)}(\gamma) +\varepsilon;\theta)-\frac{\lfloor \gamma N \rfloor}{N}\bigg)^2\bigg)\\
&\qquad \stackrel{(e)}{\leq} \exp\left[-2 N \left(F(q^{(\theta)}(\gamma) +\varepsilon;\theta)-\gamma\right)^2\right],
\end{split}
\end{equation*}
where $(c)$ follows from~\eqref{eq: observe_quantile_charac}, $(d)$ follows from Hoeffding's inequality~\cite{hoeffding}, and $(e)$ follows from the inequality $\lfloor \gamma N \rfloor < \gamma N$. Note that $F(q^{(\theta)}(\gamma) +\varepsilon;\theta) - \gamma >0$ due to Assumption~\ref{assumption: F_diff}, according to which $F(\cdot;\theta)$ has connected support. Similarly, \(\mathbb{P}({y_{(\lfloor \gamma N \rfloor)}^{(\theta)} - q^{(\theta)}(\gamma)} < -\varepsilon) < \exp[-2 N (F(q^{(\theta)}(\gamma) +\varepsilon;\theta) 
        -\gamma)^2] \)
by applying Hoeffding's inequality in the opposite direction. Combining this with~\eqref{eq: Am_c}, we see that $\mathbb{P}(A_{k, \varepsilon}^c) \to 0$ as $k \to \infty$, which means that $\mathbb{P}(A_{k, \varepsilon}) \to 1$. \hfill $\blacksquare$} 

\subsection{Proof of Theorem~\ref{thm: Convergence of TS}} \label{appendix: proof_thm_1} 
The limit can be established as follows:
\begin{equation*}
\begin{split}
&\lim_{m \to \infty} \lim_{N \to \infty} \hat{\boldsymbol{\beta}} \\
&\quad \stackrel{(b)}{=}  \lim_{m \to \infty} \lim_{N \to \infty} \bigg[\sum_{i=1}^{m} \ve{J}(\ve{h}_N(\boldsymbol{y}^{(i)}))  \ve{J}^{\top}(\ve{h}_N(\boldsymbol{y}^{(i)}))\bigg]^{-1}\\
&\quad\quad \quad \quad  \quad \quad \quad  \cdot \sum_{i=1}^{m} \ve{J}(\ve{h}_N(\boldsymbol{y}^{(i)})) \tilde{\theta}_i \\
&\quad \stackrel{(c)}{=} \lim_{m \to \infty}  
\bigg[\frac{1}{m}\sum_{i=1}^{m} \ve{J}(\ve{h}^{(i)}(\boldsymbol{\gamma}_n)) \ve{J}^{\top}(\ve{h}^{(i)}(\boldsymbol{\gamma}_n))\bigg]^{-1}\\
&\quad \quad \quad \quad \quad \cdot  \frac{1}{m} \sum_{i=1}^{m} \ve{J}(\ve{h}^{(i)}(\boldsymbol{\gamma}_n)) \tilde{\theta}_i  \\
&\quad \stackrel{(d)}{=}\mathbb{E}_{\theta}[\ve{J}(\ve{h}^{(\theta)}(\boldsymbol{\gamma}_n)) \ve{J}^{\top}(\ve{h}^{(\theta)}(\boldsymbol{\gamma}_n))]^{-1}\mathbb{E}_{\theta}[\theta \ve{J}(\ve{h}^{(\theta)}(\boldsymbol{\gamma}_n))]\\
&\quad = \bar{\boldsymbol{\beta}}.
\end{split}
\end{equation*}
Here, $(b)$ follows from~\eqref{eq: beta_N_M}, $(c)$ follows from Lemmas~\ref{lemma: quantile convergence} and~\ref{lemma: a.s. converegence continuous function}, and $(d)$ follows from strong law of large numbers~\cite{ferguson2017course}.  \hfill $\blacksquare$

\subsection{Proof of Theorem~\ref{thm: strong consistency of TS}} \label{appendix: proof_thm2} 
We have that
\begin{equation*}
\begin{split}
&\lim_{N \to \infty} \hat{\theta}_{0} = \lim_{N  \to \infty} \bar{\boldsymbol{\beta}}^{\top} \ve{J}(\ve{h}_N(\boldsymbol{y}_{N}^{(0)}))\\
&\qquad\qquad \stackrel{(a)}{=} \lim_{N \to \infty} \mathbb{E}_{\theta}[\theta \ve{J}^{\top}(\ve{h}^{(\theta)}(\boldsymbol{\gamma}_n)) ]\\
&\qquad\qquad \quad \quad \quad \quad \,\, \cdot \mathbb{E}_{\theta}[\ve{J}(\ve{h}^{(\theta)}(\boldsymbol{\gamma}_n)) \ve{J}^{\top}(\ve{h}^{(\theta)}(\boldsymbol{\gamma}_n))]^{-1} \\
&\qquad\qquad \quad \quad \quad \quad \quad  \, \cdot \ve{J}(\ve{h}_N(\boldsymbol{y}_{N}^{(0)}))\\
&\qquad\qquad \stackrel{(b)}{=} \lim_{N \to \infty} \mathbb{E}_{\theta}[{{\boldsymbol{\beta}}^*}^{\top} \ve{J}(\ve{h}^{(\theta)}(\boldsymbol{\gamma}_n)) \ve{J}^{\top}(\ve{h}^{(\theta)}(\boldsymbol{\gamma}_n))]\\
&\qquad\qquad \quad \quad \quad \quad \quad \cdot \mathbb{E}_{\theta}[\ve{J}(\ve{h}^{(\theta)}(\boldsymbol{\gamma}_n)) \ve{J}^{\top}(\ve{h}^{(\theta)}(\boldsymbol{\gamma}_n))]^{-1}\\
&\qquad\qquad \quad \quad \quad \quad \quad \cdot \ve{J}(\ve{h}_N(\boldsymbol{y}_N^{(0)}))\\
&\qquad\qquad \stackrel{(c)}{=} \lim_{N \to \infty} {\boldsymbol{\beta}^*}^{\top} \ve{J}(\ve{h}_N(\boldsymbol{y}_{N}^{(0)}))\\
&\qquad\qquad = {\boldsymbol{\beta}^*}^{\top} \lim_{N \to \infty} \ve{J}(\ve{h}_N(\boldsymbol{y}_N^{(0)}))\\
&\qquad\qquad = {\boldsymbol{\beta}^*}^{\top} \ve{J}(\ve{h}^{(0)}(\boldsymbol{\gamma})) = \theta_0,
\end{split}
\end{equation*}
where $(a)$ follows from~\eqref{eq: bar_beta}, whereas $(b)$ and $(c)$ follow from Assumptions~\ref{assumption: identifiability} and~\ref{assumption: invertibility} respectively.
Therefore, $\lim_{N \to \infty} \hat{\theta}_{0} = \theta_0$ a.s. \hfill $\blacksquare$

\subsection{Proof of Theorem~\ref{thm: Asymptotic normality}} \label{appendix: proof_thm_3} 
Recall that $\hat{\theta}_{0} = \bar{\theta}_{\text{TS}}(\ve{h}_N(\boldsymbol{y}^{(0)}_{N})) =\bar{\boldsymbol{\beta}}^{\top} \ve{J}(\ve{h}_N(\boldsymbol{y}_N^{(0)}))$. From~\eqref{eq: Asdist of quantiles}, we know that
\begin{equation} \label{eq: asdist test sample quantiles}
     \sqrt{N}(\ve{h}_N(\boldsymbol{y}_N^{(0)}) - \ve{h}^{(0)}(\boldsymbol{\gamma}_n)) \stackrel{d}{\to} \mathcal{N}(\ve{0},\ve{\Sigma}),
\end{equation}
where $\ve{\Sigma}$ is obtained by substituting $\theta$ by $\theta_0$ in~\eqref{eq:Ascov matrix quantiles}.

Using Lemma~\ref{lemma: Delta method} with $b=1/2$, $N = N$, and $g(\ve{h}_N(\boldsymbol{y}_{N}^{(0)})) =
\bar{\boldsymbol{\beta}}^{\top} \ve{J}(\ve{h}_N(\boldsymbol{y}_{N}^{(0)}))$ in~\eqref{eq: asdist test sample quantiles}, we finally obtain
\begin{multline}
\sqrt{N}(\bar{\boldsymbol{\beta}}^{\top} \ve{J}(\ve{h}_{N}(\boldsymbol{y}_{N})) - \bar{\boldsymbol{\beta}}^{\top} \ve{J}(\ve{h}^{(0)}(\boldsymbol{\gamma}_n)) \stackrel{d}{\to} \mathcal{N}(0, \Sigma_{\text{TS}}),
\end{multline}
where $\Sigma_{\text{TS}}=\bar{\boldsymbol{\beta}}^{\top} \nabla \ve{J}(\ve{h}^{(0)}) \ve{\Sigma} \nabla \ve{J}(\ve{h}^{(0)})^{\top} \bar{\boldsymbol{\beta}}$ is the asymptotic covariance of the TS estimator, and $\nabla \ve{J}(\ve{h}^{(0)}) \in \mathbb{R}^{m \times n}$ is the Jacobian matrix of $\ve{J}$ evaluated at $\ve{h}^{(0)}(\boldsymbol{\gamma}_n)$.$\,\quad \quad \quad \quad \quad \qquad \qquad \qquad \qquad \qquad \qquad \qquad \blacksquare$
{
\section{Comparative experiment: TS vs.\ EKF and PEM} \label{app:comparison}
 We consider a controlled‑dynamical‑system example from~\cite{garatti2013new}, where the true system $\mathcal{M}(\theta_0)$ is driven by a stochastic input and has state update and output equations
\begin{equation*}
\begin{aligned}
x^{(1)}_{k+1} &= \tfrac{1}{2} x^{(1)}_k + u_k + v^{(11)}_k,\\
x^{(2)}_{k+1} &= (1 - \theta_0^2)\,\sin\!\bigl(50\theta_0^2\bigr) x^{(2)}_k - \theta_0\, x^{(2)}_k + \tfrac{\theta_0}{1+\theta_0^2}\,u_k + v^{(12)}_k,\\
y_k &= x^{(2)}_k + v^{(2)}_k,
\end{aligned}
\end{equation*}
where $v^{(11)}_k\sim\mathcal{N}(0,0.9)$, $v^{(12)}_k\sim\mathcal{N}(0,0.1)$, $v^{(2)}_k\sim\mathcal{N}(0,0.01)$, and $u_k\sim\mathcal{N}(0,1)$.  The parameter $\theta_0\in[-0.9,0.9]$ must be estimated.  

The TS estimator uses an ARX($5,5$) model for compression, computing coefficients via least squares, followed by a two-layer neural network with ReLU activations. We compare TS against:
\begin{itemize}
    \item \textbf{Extended Kalman Filter (EKF)}: The system is reformulated with an additional state \(x^{(3)}_k = \theta\):
    \begin{equation} \label{eq: ekf}
    \begin{aligned}
    x^{(1)}_{k+1} &= \frac{1}{2} x^{(1)}_k + u_k + v^{(11)}_k, \\
    x^{(2)}_{k+1} &= (1 - (x_k^{(3)})^2) \sin(50 (x_k^{(3)})^2) x^{(2)}_k - \\
    &\quad \qquad \qquad \qquad x_k^{(3)} x^{(2)}_k + \frac{x_k^{(3)}}{1 + (x_k^{(3)})^2} u_k + v^{(12)}_k, \\
    x_{k+1}^{(3)} &= x_k^{(3)} + w_k, \\
    y_k &= x^{(2)}_k + v^{(2)}_k,
    \end{aligned}
    \end{equation}
    where \(w_k \sim \mathcal{N}(0, 10^{-6})\). We test two initial error covariance matrices, denoted as $P(0)$, corresponding to the estimators EKF\_LARGE and EKF\_SMALL:
    \begin{itemize}
        \item EKF\_LARGE: \( P(0) = \begin{pmatrix} 0.1 & 0 & 0 \\ 0 & 0.1 & 0 \\ 0 & 0 & 0.5 \end{pmatrix} \).
        \item EKF\_SMALL: \( P(0) = \begin{pmatrix} 0.1 & 0 & 0 \\ 0 & 0.1 & 0 \\ 0 & 0 & 0.01 \end{pmatrix} \).
    \end{itemize}
    \item \textbf{PEM}: Minimizes \(\sum_{k=1}^N (y_k - \hat{y}_{k-1|k; \theta})^2\), with \(\hat{y}_{k-1|k; \theta} = x_{k-1}^{(2)}\). We test \( n_{\text{init}} = \{1, 5, 10\} \) random initializations.
\end{itemize}

\begin{figure}[h]
\centering
\includegraphics[width=0.9\columnwidth]{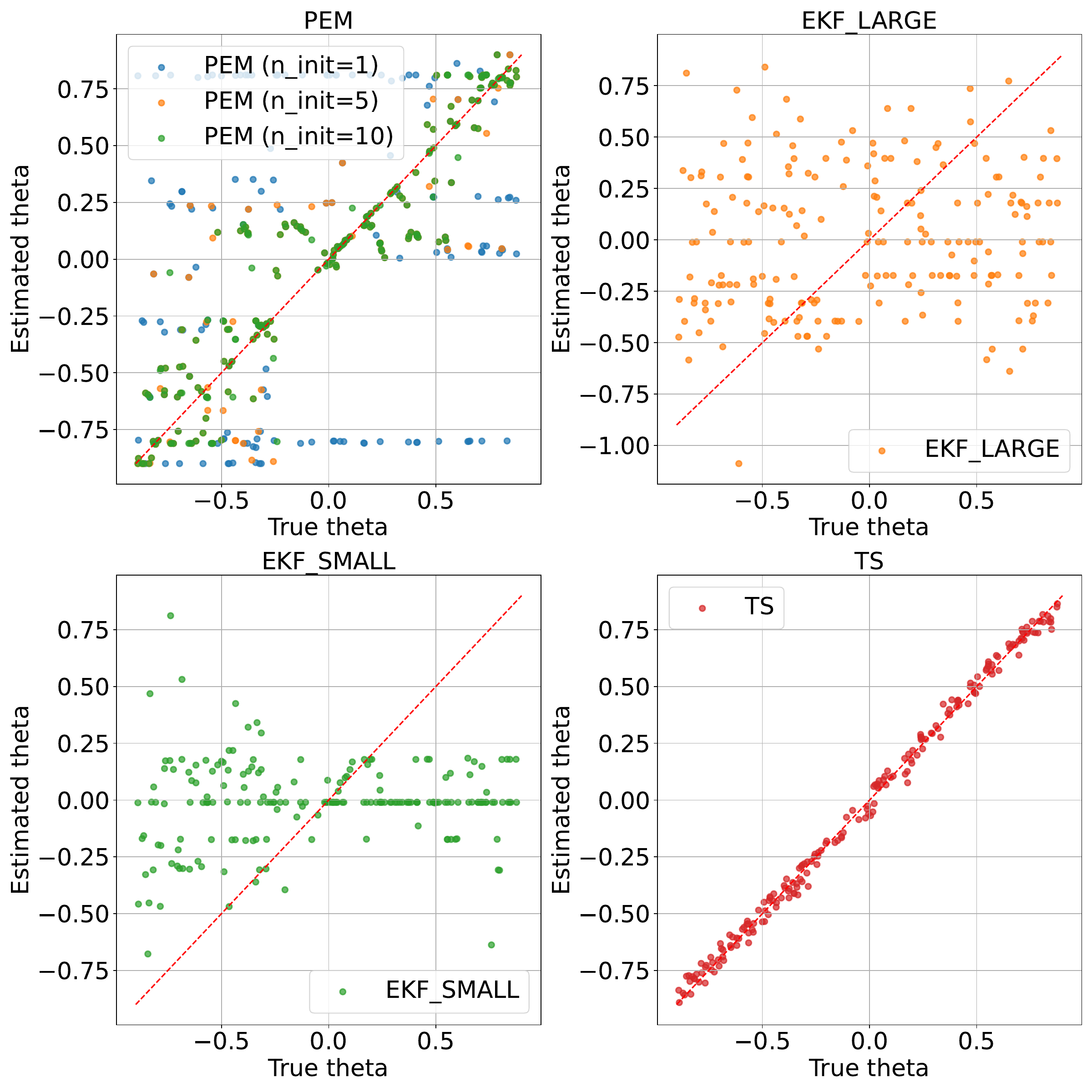}
\caption{Estimated vs.\ true parameter over Monte Carlo runs. Proximity to the diagonal ($y{=}\,x$) indicates accuracy. TS clusters tightly along the diagonal (initialization-free), whereas EKF/PEM show larger dispersion.}
\label{fig: point_estimates_comparison}
\end{figure}

\begin{figure}[h]
\centering
\includegraphics[width=0.9\columnwidth]{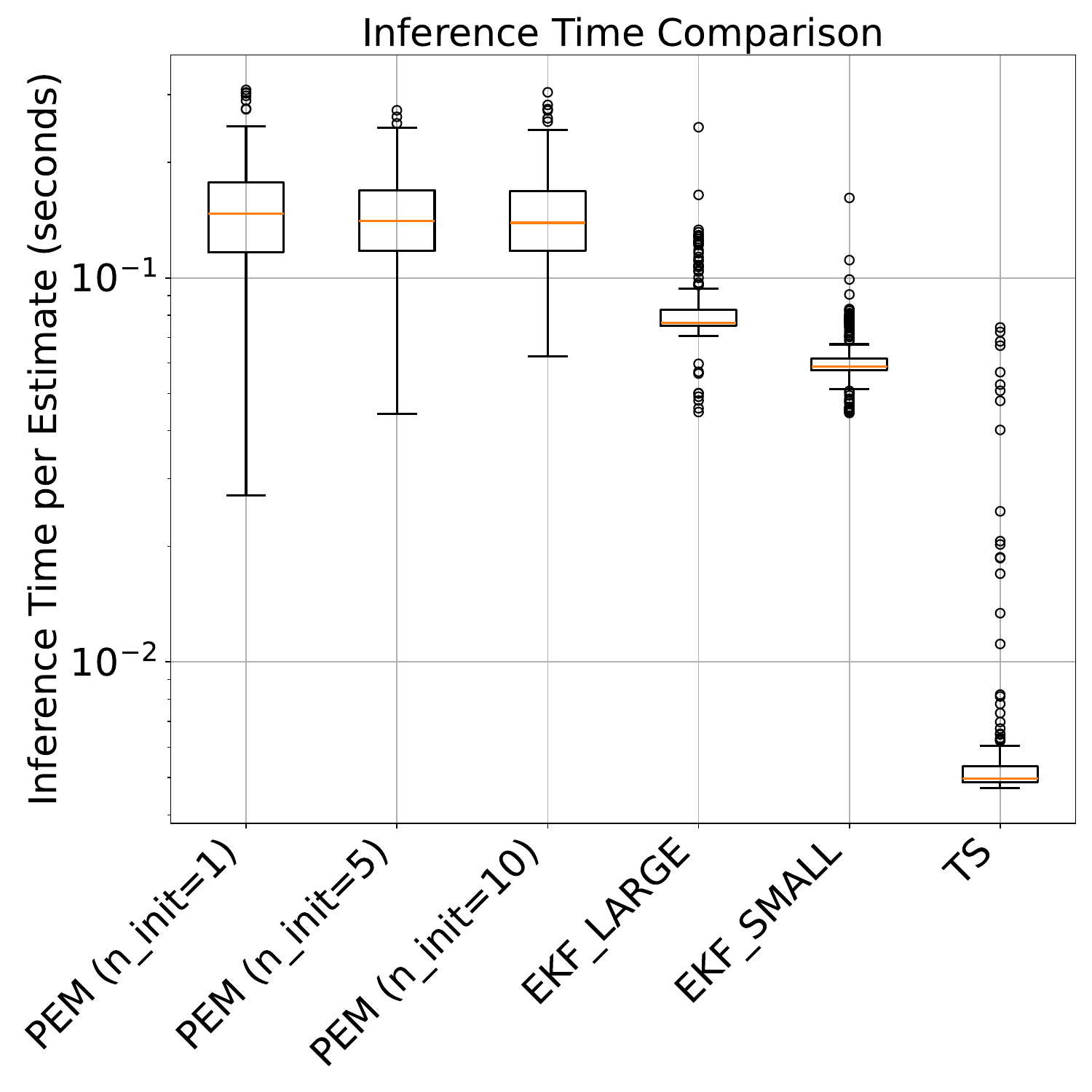}
\caption{Per-instance inference times (box plots: median line, box=IQR, whiskers=1.5\,IQR, dots=outliers). TS has the smallest median and spread, indicating faster inference.}
\label{fig: inference_time_comparison}
\end{figure}

Figure~\ref{fig: point_estimates_comparison} shows that TS produces estimates that align closely with the $y{=}\,x$ reference (\emph{i.e.}, an almost linear relation with slope~$\approx 1$), indicating high accuracy. In contrast, EKF\_LARGE, EKF\_SMALL, and PEM display off-diagonal scatter, consistent with sensitivity to initial conditions. 
Figure~\ref{fig: inference_time_comparison} summarizes the distribution of per-instance inference times: TS exhibits the smallest median and variability, indicating faster inference, highlighting its offline-trained, single-pass evaluation advantage, a benefit not explicitly quantified in \cite{garatti2013new}.
}

\bibliographystyle{plain}        
\bibliography{ref}           

\newcommand{\noop}[1]{}
\begin{thebibliography}{10}

\bibitem{caines2018linear}
P.~E. Caines.
\newblock {\em Linear stochastic systems}.
\newblock SIAM, 2018.

\bibitem{dettu2024data}
F.~Dett{\`u}, B.~Lakshminarayanan, S.~Formentin, and C.~R. Rojas.
\newblock From data to control: a two-stage simulation-based approach.
\newblock In {\em Proceedings of the 2024 European Control Conference (ECC)},
  pages 3428--3433, 2024.

\bibitem{Diskin}
T.~Diskin, Y.~C. Eldar, and A.~Wiesel.
\newblock Learning to estimate without bias.
\newblock {\em IEEE Transactions on Signal Processing}, 71:2162--2171, 2023.

\bibitem{ferguson2017course}
T.~S. Ferguson.
\newblock {\em A Course in Large Sample Theory}.
\newblock Routledge, 2017.

\bibitem{forgione2023system}
M.~Forgione, F.~Pura, and D.~Piga.
\newblock From system models to class models: An in-context learning paradigm.
\newblock {\em IEEE Control Systems Letters}, 7:3513--3518, 2023.

\bibitem{garatti2008estimation}
S.~Garatti and S.~Bittanti.
\newblock Estimation of white-box model parameters via artificial data
  generation: a two-stage approach.
\newblock {\em IFAC Proceedings Volumes}, 41(2):11409--11414, 2008.

\bibitem{garatti2013new}
S.~Garatti and S.~Bittanti.
\newblock A new paradigm for parameter estimation in system modeling.
\newblock {\em International Journal of Adaptive Control and Signal
  Processing}, 27(8):667--687, 2013.

\bibitem{GHOSH2024111327}
A.~Ghosh, M.~Abdalmoaty, S.~Chatterjee, and H.~Hjalmarsson.
\newblock {DeepBayes}--{An} estimator for parameter estimation in stochastic
  nonlinear dynamical models.
\newblock {\em Automatica}, 159:111327, 2024.

\bibitem{hoeffding}
W.~Hoeffding.
\newblock Probability inequalities for sums of bounded random variables.
\newblock {\em Journal of the American Statistical Association},
  58(301):13--30, 1963.

\bibitem{BLCRCDC2022}
B.~Lakshminarayanan and C.~R. Rojas.
\newblock A statistical decision-theoretical perspective on the two-stage
  approach to parameter estimation.
\newblock In {\em Proceedings of the 61st IEEE Conference on Decision and
  Control (CDC)}, pages 5369--5374, 2022.

\bibitem{lakshminarayanan2023minimax}
B.~Lakshminarayanan and C.~R. Rojas.
\newblock Minimax two-stage gradient boosting for parameter estimation.
\newblock In {\em Proceedings of the 62nd IEEE Conference on Decision and
  Control (CDC)}, pages 1189--1194, 2023.

\bibitem{LJUNG1976121}
L.~Ljung.
\newblock On the consistency of prediction error identification methods.
\newblock In R.~K. Mehra and D.~G. Lainiotis, editors, {\em System
  Identification Advances and Case Studies}, pages 121--164. Elsevier, 1976.

\bibitem{Ljung:99}
L.~Ljung.
\newblock {\em System Identification: Theory for the User, 2nd Edition}.
\newblock Prentice Hall, 1999.

\bibitem{mohri2008rademacher}
M.~Mohri and A.~Rostamizadeh.
\newblock Rademacher complexity bounds for non-iid processes.
\newblock {\em Advances in neural information processing systems}, 21, 2008.

\bibitem{piga2024syntheticdatagenerationidentification}
D.~Piga, M.~Rufolo, G.~Maroni, M.~Mejari, and M.~Forgione.
\newblock Synthetic data generation for system identification: leveraging
  knowledge transfer from similar systems.
\newblock {\em arXiv: 2403.05164}, 2024.

\bibitem{van2000asymptotic}
A.~W. Van~der Vaart.
\newblock {\em Asymptotic Statistics}.
\newblock Cambridge University Press, 2000.

\bibitem{vanderhorn2021digital}
E.~VanDerHorn and S.~Mahadevan.
\newblock Digital twin: Generalization, characterization and implementation.
\newblock {\em Decision support systems}, 145:113524, 2021.

\bibitem{vidyasagar2013learning}
M.~Vidyasagar.
\newblock {\em Learning and Generalisation: With Applications to Neural
  Networks}.
\newblock Springer, 2013.

\end{thebibliography}



\end{document}